\documentclass
[aps,prb,endfloat,twocolumn,showpacs,preprintnumbers,amsmath,amssymb]{revtex4}
\usepackage{natbib}
\usepackage[dvips]{graphicx}
\usepackage{wrapfig}
\usepackage{dcolumn}
\usepackage{bm}
\usepackage{times}
\usepackage{mathptmx}
\usepackage{textcomp}
\usepackage[usenames]{color}
\usepackage{ulem}
\def\geqap{\,\raise 2pt \hbox{$>\kern-11pt \lower 5pt \hbox{$\sim$}$}\,}
\def\leqap{\,\raise 2pt \hbox{$<\kern-10pt \lower 5pt \hbox{$\sim$}$}\,}
\def\rfeo{$R$Fe$_2$O$_4$ \ }
\def\fetwo{Fe$^{2+}$ }
\def\fethr{Fe$^{3+}$ }
\begin{document}
%
%
\title{Doubly degenerate orbital system in honeycomb lattice: 
\\ implication of orbital state  in layered iron oxide}

\author{J. Nasu, A. Nagano$^{\ast}$, M. Naka and S. Ishihara} 
\affiliation{Department of Physics, Tohoku University, Sendai 980-8578, Japan}
\date{\today}
%
\begin{abstract}
We study a doubly-degenerate orbital model on a honeycomb lattice. 
This is a model for orbital states in multiferroic layered iron oxides. 
The classical and quantum models are analyzed by spin-wave approximation, Monte-Carlo 
simulation and Lanczos method. 
A macroscopic number of degeneracy exists in the classical ground state. 
In the classical model, a peak in the specific heat appears at a temperature 
which is much lower than the mean-field ordering one. 
Below this temperature, 
angle of orbital pseudo-spin is fixed, but 
conventional orbital orders are not suggested. 
The degeneracy in the ground state is partially lifted by thermal fluctuation. 
We suggest a role of zero-dimensional fluctuation in hexagons on a low-temperature orbital structure.
Lifting of the degeneracy also occurs at zero temperature due to the quantum zero-point fluctuation. 
We show that the ground-state wave function is well represented by a linear combination 
of the states where a honeycomb lattice is covered by nearest-neighboring pairs of orbitals with the minimum bond energy.  
\end{abstract}
]
%
%

\maketitle
\section{introdcution}
\label{sec:intro}

Orbital degree of freedom and its interplay with spin and charge degrees are 
one of the recent attractive themes in condensed matter physics.~\cite{maekawa_rev,tokura} 
Orbital represents an anisotropic shape of the electronic wave function. 
In a molecule, this degree of freedom is quenched by the Jahn-Teller effect, and/or 
a formation of the chemical bond along a specific bond direction. 
On the contrary, in a solid crystal, some equivalent bonds coexist. 
One alignment of orbitals dose not fully satisfy the minimum-energy configuration for all equivalent bonds. 
This is a certain kind of frustration subsisting intrinsically in a solid crystal with orbital degeneracy. 
This frustrating and directional character for the orbital  
provides a wide variety of exotic phenomena in transition-metal compounds near a Mott insulating state.  

For orbital degenerate systems under strong electron correlation, 
a number of theoretical investigations have been done for more than one decade. 
One of the well known and examined orbital models is  
the so-called three-dimensional $e_g$ orbital model.~\cite{kugel82,ishihara97,feiner97} 
This is proposed as a model for orbital state in LaMnO$_3$ and KCuF$_3$ with the perovskite crystal structure. 
The doubly degenerate $e_g$ orbitals, $d_{x^2-y^2}$ and $d_{3z^2-r^2}$, are 
represented by the pseudo-spin (PS) operator ${\bf T}$ with magnitude of 1/2 
and are located on a simple cubic lattice. 
The model Hamiltonian is given by 
\begin{eqnarray}
{\cal H}_{e_g}=J \sum_{i } \left (
\tau^{x}_i \tau^{x}_{i+{\bf e}_x}
+\tau^{y}_i \tau^{y}_{i+{\bf e}_y}
+\tau^{z}_i \tau^{z}_{i+{\bf e}_z}
 \right  )  .   
 \label{eq:hameg}
\end{eqnarray}
Here, a vector 
${\bf e}_\eta$ for $\eta=(x,\ y,\ z)$ connects the nearest neighboring (NN) sites, 
and $\tau_i^\eta$ is a linear combination of the PS operator defined by 
$\tau^\eta_i=- \sin (2\pi n_\eta/3) T^z_i + \cos (2\pi n_\eta/3)  T^x_i$ 
with a factor $(n_x, n_y, n_z)=(1, 2, 3)$. 
This model is derived by the perturbational procedure from the extended Hubbard Hamiltonian 
with neglecting spin degree of freedom. 
The $\eta$ dependence of the interaction implies the frustrating and directional character. 
As seen in frustrated magnets, there is a macroscopic number of degeneracy in the classical ground state. 
This degeneracy is lifted by thermal fluctuation in finite temperatures 
and by quantum zero-point fluctuation.~\cite{nussinov04,biskup05,kubo02,tanaka05} 
As results, a staggered-type long-range orbital order is realized. 

Doubly-degenerate orbital model on a honeycomb lattice, studied in the present paper, 
is one of the orbital models with the frustrating and directional interaction. 
Orbital degree of freedom represented by the PS operator  
is located on a two-dimensional honeycomb lattice [see Fig.~\ref{fig:lattice}]. 
An explicit form of the Hamiltonian 
is given in Eq.~(\ref{eq:ham}), which is introduced in more detail in Sect.~\ref{sect:model}. 
This model looks similar to the $e_g$ orbital model in Eq.~(\ref{eq:hameg}); 
the NN three-bond directions in a honeycomb lattice, $\alpha$, $\beta$, and $\gamma$, 
correspond to the Cartesian coordinates in a cubic lattice. 
Thus, a similar kind of frustrating character for orbital configuration is expected. 
However, in general, stability of an orbital state is 
extremely sensitive to symmetry and dimension of a crystal lattice. 
It is nontrivial whether a conventional long-range order is realized or not, 
in the same type of interaction, but in the different crystal lattice. 
From a viewpoint of substantial materials, 
the honeycomb-lattice orbital model is proposed as an orbital model in     
a multiferroic layered iron oxide $R$Fe$_2$O$_4$ ($R$=Lu, Y, Yb).~\cite{nagano07a,nagano07} 
This is a mixed valence compound where equal amount of Fe$^{2+}$ and Fe$^{3+}$ coexists 
in a pair of triangular lattice.~\cite{kimizuka90,ikeda05,yamada00,akimitsu79,shiratori92} 
A Fe$^{2+}$ ion with $d^6$ configuration has the doubly degenerate orbital degree of freedom. 
In the low-temperature charge and spin ordered phase, a Fe$^{2+}$ sublattice forms a honeycomb lattice, 
and the orbital state is mapped onto a honeycomb lattice model. 
This will be introduced later in more detail. 
From different view point, 
this orbital model is proposed recently in study of the optical lattice.~\cite{wu07,wu08,zhao08}

In this paper, we study the ground-state and finite-temperature properties in the doubly-degenerate orbital model on a honeycomb lattice. We analyze the classical and quantum models by the Monte-Carlo (MC) and Lanczos methods, respectively,  
as well as the spin-wave approximation. There is a number of the degenerate classical ground states as well as the $e_g$ orbital model. In the classical model, at a certain temperature which is much lower than the mean-field ordering temperature, a peak in the specific heat appears. Below this temperature, the PS angles are fixed at one of $\pi n/3$ with an integer number $n$. 
The degeneracy is partially lifted below this temperature due to thermal fluctuation, but the conventional long-range orders are not suggested from the two-body correlation functions for PS. 
This degeneracy is also lifted by the quantum zero-point fluctuation. 
The ground-state wave-function is well reproduced by a linear combination of the states that a honeycomb lattice is covered by dimer pairs of the NN PS configurations which satisfy the minimum bond energy. 

In Sect.~\ref{sect:model}, we define the Hamiltonian of the honeycomb lattice orbital model, 
and introduce implication of the orbital state in layered iron oxides. 
Results in the classical and quantum models are presented in Sects.~\ref{sect:classical}  
and \ref{sect:quantum}, respectively. 
Section \ref{sect:conclusion} is devoted to the discussion and summary. 
Preliminary results  
have been published in Refs.~\onlinecite{nagano07a} and \onlinecite{nagano07}. 
Relation to the layered iron oxides is briefly introduced in Ref.~\onlinecite{naka08}. 

\section{Model}
\subsection{Model Hamiltonian}
\label{sect:model}

\begin{figure}
\includegraphics[width=0.9\columnwidth]{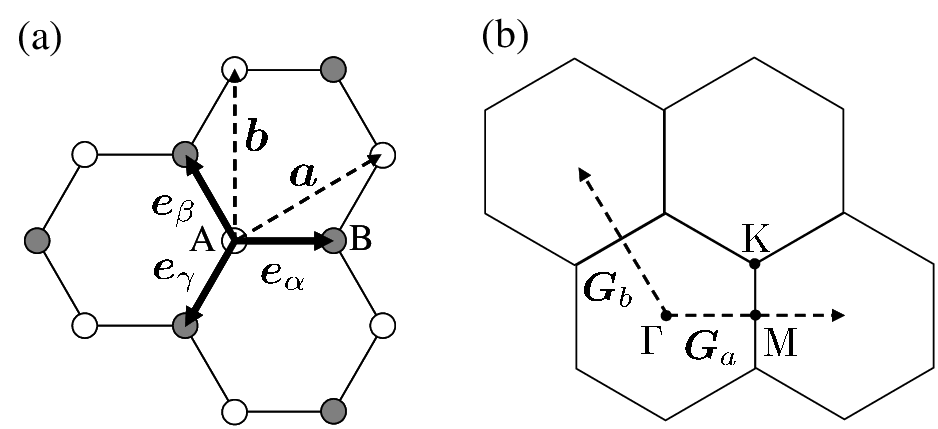}
\caption{ 
(a) A honeycomb lattice structure and sublattices A and B. 
Bold arrows represent vectors connecting NN two sites.  
(b) Brillouin zone and reciprocal lattice vectors for a honeycomb lattice. 
} 
\label{fig:lattice}
\end{figure}
\begin{figure}
\includegraphics[width=0.45\columnwidth]{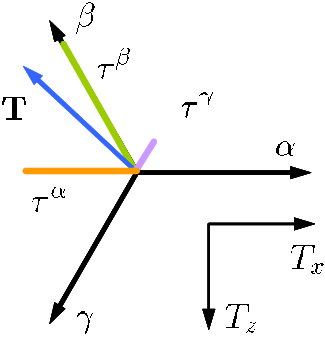}
\caption{ 
Pseudo-spin operator ${\bf T}$, and its projection components $\tau^\eta$ along the three-bond directions.  
} 
\label{fig:tau}
\end{figure}

We start with the model Hamiltonian for the doubly degenerate orbitals, 
denoted by $a$ and $b$, defined in a honeycomb lattice.
This is represented by the pseudo-spin operator 
with magnitude of $1/2$:  
\begin{equation}
{\bf T}_i=\frac{1}{2} \sum_{t t' s} 
d_{i t s }^\dagger {\bf \sigma}_{t t'} d_{i t' s} , 
\label{eq:ps}
\end{equation}
where $d_{i t s}$ is the electron annihilation operator with orbital $t(=a, b)$, 
spin $s(=\uparrow, \downarrow)$ at site $i$, and $\bf \sigma$ are the Pauli matrices. 
For the three-kinds of NN bonds, $\eta=(\alpha, \beta, \gamma)$, in a honeycomb lattice (see Fig.~\ref{fig:lattice}), 
we introduce new PS operator as 
\begin{eqnarray}
\tau^\eta_i=- \sin \left (\frac{2\pi n_\eta}{3} \right ) T^z_i + \cos \left (\frac{2\pi n_\eta}{3} \right ) T^x_i . 
\end{eqnarray}
A numerical factor $n_\eta$ is defined as $(n_\alpha, n_\beta, n_\gamma)=(0,1,2)$. 
When we define the pseudo-spin coordinate as shown in Fig.~\ref{fig:tau}, 
the operator $\tau^\eta_i$ represents a projection component of ${\bf T}_i$ on 
the $\eta$ bond direction. 
The model Hamiltonian  studied in the present paper is, 
\begin{eqnarray}
{\cal H}=-J \sum_{i \in A} \left (
\tau^{\alpha}_i \tau^{\alpha}_{i+{\bf e}_\alpha}
+\tau^{\beta}_i \tau^{\beta}_{i+{\bf e}_\beta}
+\tau^{\gamma}_i \tau^{\gamma}_{i+{\bf e}_\gamma}
 \right  )  ,  
 \label{eq:ham}
\end{eqnarray}
where ${\bf e}_\eta$ is a vector connecting the NN sites along the direction $\eta$, 
$\sum_{i \in {\rm A}}$ represents a sum of sites on the sublattiece A [see Fig.~\ref{fig:lattice}(a)], 
and $J$ is the exchange constant. 
Although $J$ is defined to be positive, 
its sign is gauged away by rotating PS's on the A sublattice with respect to $T^y$. 
This Hamiltonian is rewritten as a following simple form 
\begin{eqnarray}
{\cal H}=\frac{J}{2} \sum_{i \in A, \eta}
\left ( \tau_i^\eta -\tau_{i+{\bf e}_\eta}^\eta \right )^2
-\frac{3}{2}J \sum_{i \in A} \left ( T_i^{x2}+T_i^{z2} \right ) . 
\label{eq:ham2}
\end{eqnarray}
The second term is $-3JN/16$, when ${\bf T_i}$ is a two-dimensional classical spin, 
and is $-3JN/8$ in the quantum-spin case. 
A total number of sites is $N$. 
This model is proposed as a orbital state for the layered iron oxide,~\cite{nagano07a,nagano07} 
as explained in Sect.~\ref{sect:rfeo} in more detail, 
and is also recently proposed in study of the optical lattice.~\cite{wu07,wu08,zhao08}
A similar orbital model in a honeycomb lattice termed the Kitaev model is recently well examined.~\cite{kitaev06,baskaran07}
Here three components of the PS operator, $T^l_i$ with $l=(x,y,z)$, instead of $\tau_i^l$ in this model, 
are concerned in the interactions along the $\alpha$, $\beta$ and $\gamma$ directions.  

\begin{figure}
\includegraphics[width=0.8 \columnwidth]{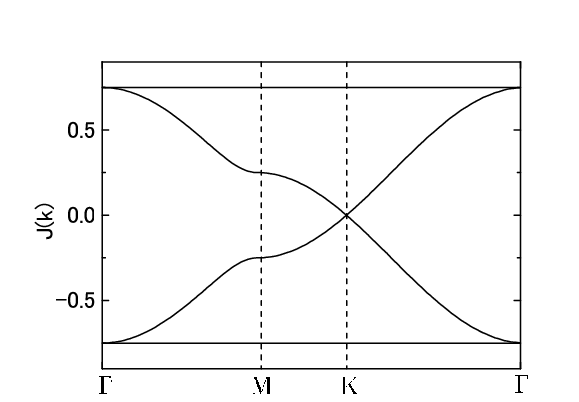}
\caption{ 
Eigen values of the orbital interaction $\widehat J({\bf k})$ in the momentum space. 
} 
\label{fig:jk}
\end{figure}
Before going to detailed analyses of the Hamiltonian, 
we briefly introduce a character in this model. 
Let introduce the Fourier transformation for the orbital PS, 
\begin{eqnarray}
{\bf T}_{\rm C} \left ({\bf k} \right)=\frac{1}{\sqrt{N/2}} 
\sum_{i \in {\rm  C}} {\bf T}_i e^{i {\bf k} \cdot {\bf r_i}} , 
\end{eqnarray}
for the sublattice C(=A, B).  
The Hamiltonian Eq.~(\ref{eq:ham}) is represented in the momentum space,~\cite{nagano07,wu07} 
shown in Fig.~\ref{fig:lattice}(b), as 
\begin{eqnarray}
{\cal H}= 
\psi^t \left ( -{\bf k}\right )
{\hat J} \left ( {\bf k} \right )
\psi \left ( {\bf k} \right ) . 
\end{eqnarray}
We introduce 
a four-component vector defined as 
\begin{equation}
\psi({\bf k})=\left [ T_{\rm A}^x({\bf k}), T_{\rm A}^z({\bf k}), T_{\rm B}^x({\bf k}), T_{\rm B}^z({\bf k}) \right ] , 
\end{equation}
and a $4 \times 4$ matrix ${\hat J}({\bf k})$. 
We obtain the eigen values of ${\hat J}({\bf k})$ are
$\pm 3J/4$ and 
$\pm J
\left [ 3 + 2\cos {\bf k} \cdot {\bf a}+2 \cos {\bf k} \cdot {\bf b} +2\cos {\bf k} \cdot ({\bf a-b}) 
\right ] ^{1/2}/4$ 
where $\bf a$ and $\bf b$ are the primitive translation vectors 
defined in Fig.~\ref{fig:lattice}. 
Numerical plot of $\hat J({\bf k})$ is presented in Fig.~\ref{fig:jk}. 
The lowest eigen value is a momentum independent flat band of $-3J/4$. 
That is, the effective dimensionality for the lowest state is zero, and, 
in the classical ground state,  
stable orbital structures are not determined uniquely due to large fluctuation. 
The second eigen value touches the lowest band at the point $\Gamma$. 

Compare the present model with the $e_g$ orbital model in a simple cubic lattice. 
The $e_g$ orbital model defined in Eq.~(\ref{eq:hameg}) shows a similar 
form with the present honeycomb lattice model in Eq.~(\ref{eq:ham}),  
when $\alpha$, $\beta$ and $\gamma$ are replaced by the Cartesian coordinates $x$, $y$ and $z$. 
The momentum representation of the orbital interaction is given by 
$\hat J({\bf k})=\pm J [ 3+\cos k_x a +\cos k_y a +\cos k_z a ]^{1/2}$
where $(k_x, k_y, k_z)$ are defined in the Brillouin zone for a simple cubic lattice.~\cite{ishihara97B}
Dispersion relation of $\hat J({\bf k})$ is flat 
along $(\pi, \pi, \pi)-(0, \pi, \pi)$ and other equivalent directions. 
Due to the flat dispersions, there is 
a macroscopic number of degeneracy in the classical ground state. 
However, this degeneracy is lifted by thermal and quantum fluctuations, 
and a staggered long-range orbital order is realized.~\cite{nussinov04,kubo02} 
This is the so-called order-by-fluctuation mechanism.  
The long-range order in the classical model is confirmed  
by the Monte-Carlo simulation; 
the two-body correlation function for PS at momentum ${\bf k}=(\pi, \pi, \pi)$ starts to increase around $T=0.17J$, 
and is saturated at its maximum value in the low temperature limit 
[see inset in Fig.~\ref{fig:corr_class}(c)].~\cite{tanaka05,tanaka08} 

\subsection{Implication of layered iron oxide}
\label{sect:rfeo}
\begin{figure}
\includegraphics[width=0.9\columnwidth]{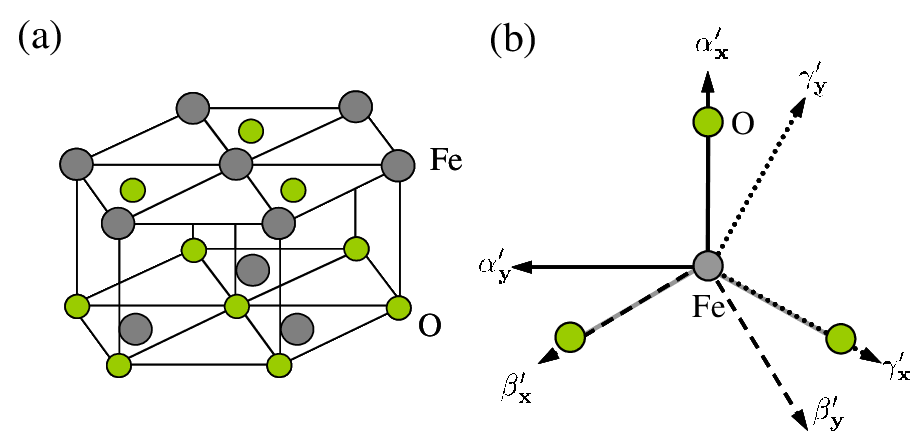}
\caption{ 
(a) A pair of triangular planes termed the W-layer, and (b) three Fe-O bond directions in a triangular lattice 
in $R$Fe$_2$O$_4$. 
} 
\label{fig:rfeo}
\end{figure}

In this section, we introduce the honeycomb lattice orbital model defined in Eq.~(\ref{eq:ham}) 
as an orbital model for multiferroic layered-iron oxides $R$Fe$_2$O$_4$. 
This is known as a multiferroic material driven by electronic charge and spin degrees of freedom.  
Electric and magnetic properties in \rfeo are dominated by 
Fe $3d$ electrons in a pair of triangular-lattice planes stacked along the $c$ axis, 
which is termed the W-layer [see Fig.~\ref{fig:rfeo}(a)]. 
A Fe ion in the W-layer is five fold coordinate with a local symmetry of D$_{\rm 3d}$. 
The five $3d$ orbitals under the crystalline field split into 
two sets of the doubly degenerate orbitals, $\{ d_{xy}, d_{x^2-y^2} \}$ with the symmetry ${\rm E'}$, 
and $\{ d_{yz}, \ d_{zx} \}$ with ${\rm E''}$, 
and the $d_{3z^2-r^2}$ orbital with ${\rm A'}$. 
We obtained by the crystalline field calculation 
that the ${\rm E'}$ orbital is the lowest. 
Since a nominal valence of the Fe ions is 2.5+, 
equal amount of Fe$^{2+}$ ($d^6$) and Fe$^{3+}$ ($d^5$) coexists. 
The five $3d$ orbitals are singly occupied in Fe$^{3+}$, 
and one of the degenerate lowest orbitals in Fe$^{2+}$ is doubly occupied. 
Thus, \fetwo has the doubly degenerate orbital degree of freedom. 
This is represented by the PS operator defined in Eq.~(\ref{eq:ps}) 
where $t$ takes $d_{xy}$ and $d_{x^2-y^2}$.  
It is convenient to introduce the three two-dimensional coordinates 
$(\eta_x', \eta_y')$ with 
$ \eta'=(\alpha', \beta', \gamma' )$ where 
the $\eta_x'$ axis is parallel to one of the NN Fe-O bonds  
as shown in Fig.~\ref{fig:rfeo}(b).  
We define, in these coordinates, linear combinations of the orbital operators: 
\begin{eqnarray}
\left ( 
\begin{array}{c}
d_{i \eta_x^{'2}-\eta_y^{'2} s} \\
d_{i \eta_x' \eta_y' s}
\end{array}
\right ) &=&
\left (  
\begin{array}{rr}
 \cos\frac{4\pi}{3}n_{\eta'} , &\sin\frac{4\pi}{3}n_{\eta'} \\
-\sin\frac{4\pi}{3}n_{\eta'}, & \cos\frac{4\pi}{3}n_{\eta'}
\end{array}  
\right ) 
\nonumber \\
&\times&
\left ( 
\begin{array}{c}
d_{i x^{2}-y^{2} s} \\
d_{i xy s}
\end{array}
\right ) , 
\end{eqnarray}
with a numerical factor $(n_{\alpha'}, n_{\beta'}, n_{\gamma'})=(0,1,2)$. 
In the NN Fe-O bond along the $\eta'_x$ axis, 
the $d_{\eta_x^{'2}-\eta_y^{'2}}$ 
and O $2p$ orbitals form the $\sigma$ bond. 
We redefine the PS operators, 
\begin{eqnarray}
\tau^{\eta_i'}=
\cos \left (\frac{2\pi}{3} n_{\eta'} \right ) T^z_i + \sin \left (\frac{2\pi}{3} n_{\eta'} \right ) T^x_i .  
\label{eq:tau2}
\end{eqnarray}
One hole occupied state in the $d_{\eta_x^{'2}-\eta_y^{'2}}$ ($d_{\eta_x' \eta_y'})$ orbital at site $i$ 
is the eigen state of $\tau^{\eta '}_i$. 

\begin{figure}[t]
\includegraphics[width=0.8\columnwidth]{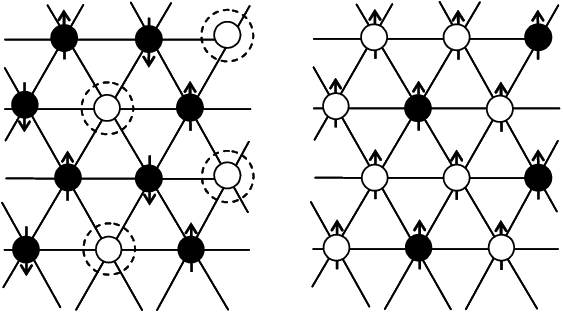}
\caption{ 
Schematic picture of the charge and spin structures in 2Fe$^{2+}$-Fe$^{3+}$ plane (right) 
and in Fe$^{2+}$-2Fe$^{3+}$ plane (left) for $R$Fe$_2$O$_4$.  
Filled and open circles represent Fe$^{3+}$ and Fe$^{2+}$, respectively. 
At sites surrounded by dotted circles, spin directions are not uniquely determined due to frustration. 
} 
\label{fig:spin3}
\end{figure}
Interaction between the orbitals is constructed from the 
electronic processes in a W-layer. 
The model Hamiltonian in low energy spin, charge and orbital states 
is derived from the extended $pd$ model by the perturbational procedure. 
The obtained Hamiltonian consists of the long-range Coulomb interactions between charges  
and the exchange interactions between NN spins and orbitals. 
We analyze numerically the Hamiltonian by the classical MC method. 
Details were presented in Refs.~\onlinecite{nagano07} and \onlinecite{naka08}.  
Obtained charge and spin ordered structure is shown in Fig.~\ref{fig:spin3}, which 
is consistent with the electron and neutron diffraction 
experiments.~\cite{yamada00,ikeda05,akimitsu79,shiratori92}  
A charge imbalance of Fe$^{2+}$ and Fe$^{3+}$ is realized between the 
triangular-lattice planes. 
That is, the electric dipole moment is caused by the charge order without inversion symmetry.~\cite{yamada00} 
In the spin structure shown in Fig.~\ref{fig:spin3}, 
\fetwo ions, which have the orbital degree of freedom,  
are surrounded by NN \fethr in the \fetwo-2\fethr plane,  
and these form a honeycomb lattice in the 2\fetwo-\fethr plane. 
The superexchange interactions in three \fetwo-\fethr bonds 
connecting \fetwo at site $i$ is proportional to 
$\sum_{\eta'}  {\tau}^{\eta'}_i$.
This is because the orbital is only active in \fetwo, 
and spin configurations in the three bonds are equivalent.  
It is easily shown from Eq.~(\ref{eq:tau2}) that this is zero. 
Therefore, 
the orbital degree of freedom in the charge and spin ordered phase is 
described by the Hamiltonian in a honeycomb lattice in the 2Fe$^{2+}$-Fe$^{3+}$ plane, 
\begin{eqnarray}
{\cal H}'=-J' \sum_{i \in {\rm A}} \left (
 \tau^{\alpha'}_i \tau^{\beta'}_{i+{\bf e}_\gamma}
+\tau^{\beta'}_i  \tau^{\gamma'}_{i+{\bf e}_\alpha}
+\tau^{\gamma'}_i \tau^{\alpha'}_{i+{\bf e}_\beta}
 \right  )  . 
 \end{eqnarray}
The exchange constant $J'(>0)$ is given by the intra-site 
Coulomb interactions and the hopping integrals. 
Then, we introduce the unitary transformation, 
\begin{eqnarray}
U=\exp \left \{ -i \left ( \frac{\pi}{6}\sum_{j \in {\rm A}} + \frac{5\pi}{6} \sum_{j \in {\rm B}} \right )  T_j^y  \right \}, 
\end{eqnarray}
which rotates PS's on sublattice A(B)
by angle $\pi/6$ $(5\pi/6)$ with respect to the $T^y$ axis. 
We show that $U^{-1} {\cal H}' U$ is identical to ${\cal H}$ in Eq.~(\ref{eq:ham}) where $J$ corresponds to $J'$. 
In addition to the exchange interaction described by this Hamiltonian, 
there may be some other factors which couple with orbital degree of freedom. 
However, this Hamiltonian is expected to provide a starting point 
to examine the low temperature orbital structure in layered iron oxides.

\section{Classical orbital state}
\label{sect:classical}
In this section, we treat the orbital pseudo-spin ${\bf T}_i$ as a classical two-dimensional vector 
with an amplitude of $1/2$. 

\begin{figure}
\includegraphics[width=0.9\columnwidth]{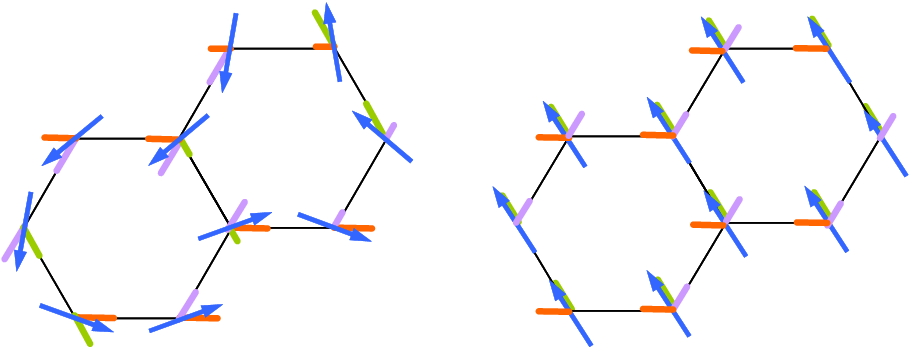}
\caption{ 
Pseudospin configurations in the ground state. 
Arrows represent directions of PS's and bold bars are for the projection components 
$\tau_i^\eta$ along the bond direction.  
A PS configuration, obtained by a uniform rotation of all PS's in right figure, is also in the ground state. 
} 
\label{fig:gs}
\end{figure}
\subsection{Orbital structure at ground state}
\label{sect:gs}

Orbital structure in the classical ground state is obtained 
from the Hamiltonian in Eq.~(\ref{eq:ham2}). 
The ground state energy is $-3J/16$, 
when the PS's satisfy the following condition in all NN bonds;~\cite{nagano07,wu08,zhao08} 
\begin{equation}
\tau_i^\eta=\tau_{i+{\bf e}_\eta}^\eta. 
\label{eq:condition}
\end{equation}
This relation implies that the projection components of PS's are equal 
with each other for all NN bonds.
There is a macroscopic number of orbital structures 
which satisfy this condition. 
Two of them are shown in Fig.~\ref{fig:gs}. 
In particular,  
uniform orbital alignments with any PS angles are in the ground state configurations. 
This kind of rotational symmetry is not expected from the Hamiltonian 
where any continuous symmetries do not exist in the PS space. 
But this is consistent with the momentum representation of the orbital interaction, $\hat J({\bf k})$; 
the second-lowest band in $\hat J({\bf k})$ touches the lowest one  
at the point $\Gamma$ as shown in Fig.~\ref{fig:jk}. 

\subsection{Spin wave analyses}

At the first step, among the degenerate uniform configurations at zero temperature, 
we turn up stable states in finite temperatures by using the spin-wave approximation.~\cite{nussinov04} 
We define the PS angle as 
$\theta_i=-\tan^{-1} (T_i^z/T_i^x)$, 
and denote an angle in the uniform configuration by $\theta^\ast$. 
A deviation from $\theta^\ast$ at site $i$ is represented by $\zeta_i(\equiv \theta_i-\theta^\ast)$. 
Within the second order of $\zeta_i$, 
the spin-wave Hamiltonian is obtained as 
\begin{eqnarray}
{\cal H}_{\rm SW}=\frac{J}{2} \sum_{i \in {\rm A} \  \eta} q_\eta \left ( \theta^\ast \right ) \left ( \zeta_{i}-\zeta_{i+{\bf e}_\eta} \right ) ^2 , 
\end{eqnarray}
where $q_\eta (\theta^\ast)=\frac{1}{4} \sin^2 [ \theta^\ast +(2\pi n_\eta )/3 ] $. 
By introducing the Fourier transform of $\zeta_i$ 
defined by 
$\zeta_{\bf k}^{\rm C}=(N/2)^{-1/2} \sum_{i \in {\rm C}} e^{i {\bf k} \cdot {\bf r_i}} \zeta_i$ for sublattice C(=A,B), 
the Hamiltonian is rewritten in a momentum space as 
\begin{eqnarray}
{\cal H}_{\rm SW}=\frac{J}{2} \sum_\eta q_\eta(\theta^\ast ) \sum_{\bf k} 
\left | \zeta^{\rm A}_{\bf k}-\zeta_{\bf k}^{\rm B} e^{-i {\bf k} \cdot {\bf e}_\eta} \right |^2 . 
\end{eqnarray}
Then, we calculate the partition function for the PS fluctuation around $\theta^\ast$. 
By introducing the two-dimensional polar coordinates 
defined by $\zeta^{\rm C}_{\bf k}=|\zeta_{\bf k}^{\rm C}| e^{i \varphi_{\bf k}^{\rm C}}$ for C=A and B, 
the partition function is obtained as 
\begin{eqnarray}
Z(\theta^\ast)&=&A \Pi_{\bf k}' \int  d|\zeta^{\rm A}_{\bf k}| d|\zeta^{\rm B}_{\bf k}| 
d\varphi^{\rm A}_{\bf k}  d\varphi^{\rm B}_{\bf k} \ 
|\zeta_{\bf k}^{\rm A}| |\zeta_{\bf k}^{\rm B}|
\nonumber \\
&\times&
\exp \left [ -\beta J  \sum_\eta q_\eta(\theta^\ast) 
\left | |\zeta^{\rm A}_{\bf k}| -|\zeta^{\rm B}_{\bf k}|  
e^{i \Delta \varphi_{\bf k}} 
e^{-i {\bf k} \cdot {\bf e}_\eta}  \right |^2 \right ] , 
\end{eqnarray}
where $A(>0)$ is the Jacobian, $\beta$ is the inverse temperature, 
$\Delta \varphi_{\bf k} =\varphi_{\bf k}^{\rm B}-\varphi_{\bf k}^{\rm A}$, 
and $\Pi_{\bf k}'$ represents a product of ${\bf k}$ in a half of the 
first Brillouin zone. 
At low temperature, 
the upper limits in the integrals for $|\zeta^{\rm A}_{\bf k}|$ and $|\zeta_{\bf k}^{\rm B}|$ are safely taken to be infinity. 
By integrating out  a variable $|\zeta^{\rm A}_{\bf k}|$, 
we obtain the following expression for the free energy, 
\begin{eqnarray}
F(\theta^\ast)&=&-\frac{1}{\beta}\log A-\frac{N}{4 \beta} \log \frac{\pi}{(\beta J)^2}
\nonumber \\
&-&\frac{1}{\beta} \sum_{\bf k}' 
f\left ( \theta^\ast,  {\bf k} \right ) , 
\end{eqnarray}
with 
\begin{eqnarray}
f\left ( \theta^\ast,  {\bf k} \right )&=&\log \int^\infty_0 d \zeta \int^{2\pi}_0 d \varphi 
\nonumber \\
&\times&
\frac{\zeta}{\left [ \sum_\eta q_\eta (\theta^\ast)  
|1-\zeta e^{i \varphi} e^{-i {\bf k} \cdot{\bf e}_\eta }|^2   \right ] ^2} , 
\end{eqnarray}
where $\sum_{\bf k}'$ represents a sum of ${\bf k}$ in a half of the first Brillouin zone. 

\begin{figure}
\includegraphics[width=\columnwidth]{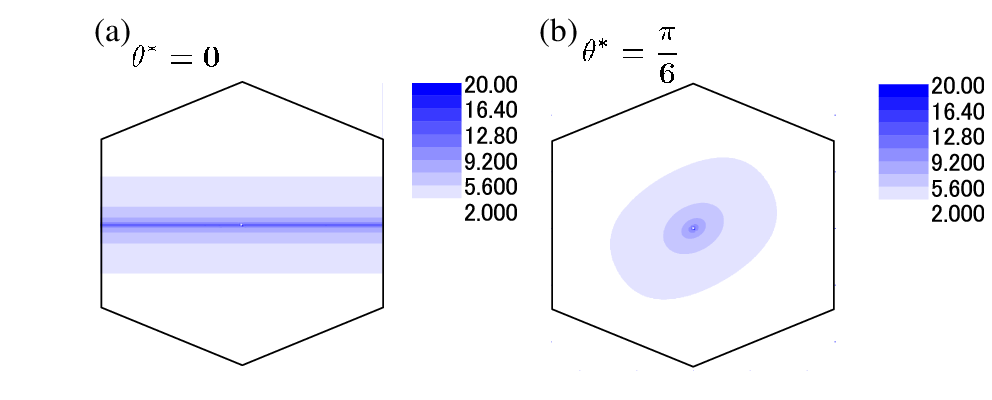}
\caption{ Contour map of the function $f(\theta^\ast, {\bf k})$ 
in the Brillouin zone for $\theta^\ast=0$ in (a), and that for $\theta^\ast=\pi/6$ in (b). } 
\label{fig:map}
\end{figure}
\begin{figure}
\includegraphics[width=\columnwidth]{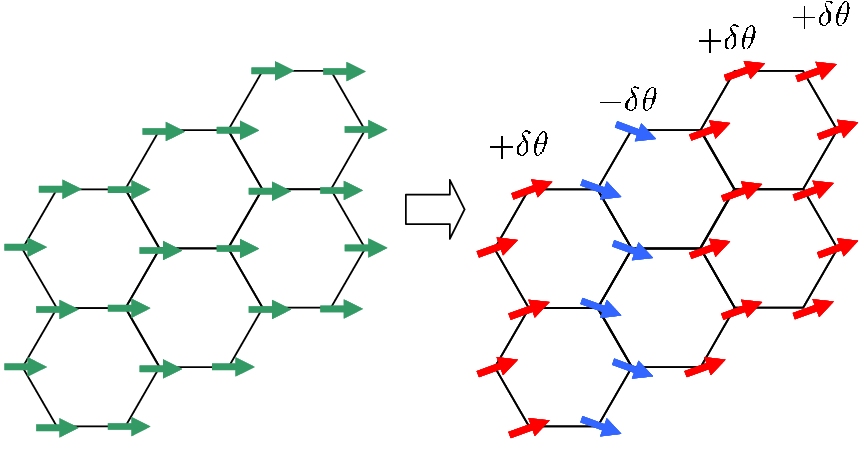}
\caption{ 
Left: PS configuration for $\theta^\ast=0$. 
Right: configuration obtained by $\pm \delta \theta $ rotations of PS's in each zigzag chain. 
} 
\label{fig:theta0}
\end{figure}
\begin{figure}
\includegraphics[width=0.8\columnwidth]{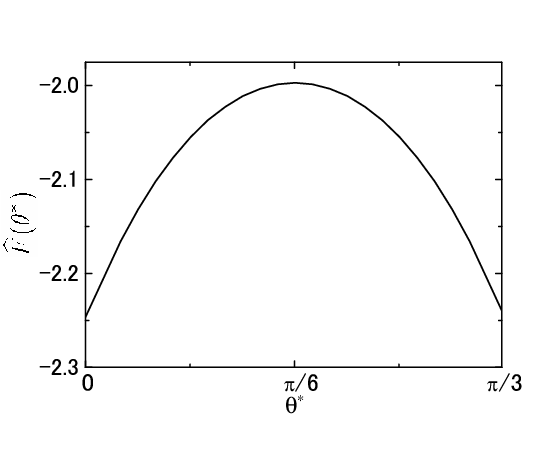}
\caption{ A part of the free energy $\widehat F(\theta^\ast)$ as a function of the PS angle $\theta^\ast$ 
obtained in the spin wave approximation. } 
\label{fig:f_theta}
\end{figure}

We numerically calculate $f(\theta^\ast, {\bf k})$. 
Contour maps of $f(\theta^\ast, {\bf k})$ for $\theta^\ast=0$ and $\pi/6$
are presented in Fig.~\ref{fig:map}. 
Results in other $\theta^\ast=2\pi n/6$ and $(2n+1)\pi/6$ with integer $n$ 
are obtained by considering the C$_6$ symmetry in $f(\theta^\ast,{\bf k})$. 
This symmetry is attributed to the fact that 
the Hamiltonian is invariant under (i) the inversion with respect to 
PS, and (ii) a combined operation of the ${\rm C}_3$ rotation for PS and that for the crystal lattice. 
In $f(\theta^\ast=0,{\bf k})$, 
a divergent behavior appears along the ${\bf G}_a$ (horizontal) axis. 
This originates from a number of low-lying PS configurations  
from the $\theta^\ast=0$ state, explained as follows. 
Start with the PS configuration with $\theta^\ast=0$ shown in Fig.~\ref{fig:theta0}, 
and focus on zigzag chains running along the ${\bf b}$ (vertical) axis. 
Rotate PS's by angle $+ \delta \theta$ or $-\delta \theta$, 
where $|\delta \theta|$ is taken to be uniform  
and their signs are chosen independently for the each zigzag chain. 
One example is shown in Fig.~\ref{fig:theta0}. 
This rotation does not change the energy, 
since the condition in Eq.~(\ref{eq:condition}) is still satisfied in all NN bonds.
On the contrary, in $\theta^\ast=\pi/6$, a 
divergent behavior in $f(\theta^\ast, {\bf k})$ is only seen at the point $\Gamma$. 
This corresponds to a uniform PS rotation. 
By integrating out the momentum ${\bf k}$ for $f(\theta^\ast, {\bf k}) $, 
we obtain the $\theta^\ast$ dependence of the free energy. 
We present, in Fig.~\ref{fig:f_theta}, 
a part of the free energy defined by  
\begin{equation}
\widehat F(\theta^\ast)=-\frac{2}{N}\sum_{\bf k}'  f(\theta^\ast, {\bf k}) . 
\end{equation}
Because of the one-dimensional fluctuation in $f(\theta^\ast, {\bf k})$, 
$\widehat F(\theta^\ast)$ takes its minima at six angles of $\theta^\ast=n \pi/3$. 
An analytical form is given as $\widehat F(\theta^\ast=n \pi/3)=-\log (16/3)-(1/2)\log \pi \approx -2.246$. 
Among the continuous uniform states, 
these six states are stabilized selectively by thermal fluctuation. 

\begin{figure}
\includegraphics[width=0.8\columnwidth]{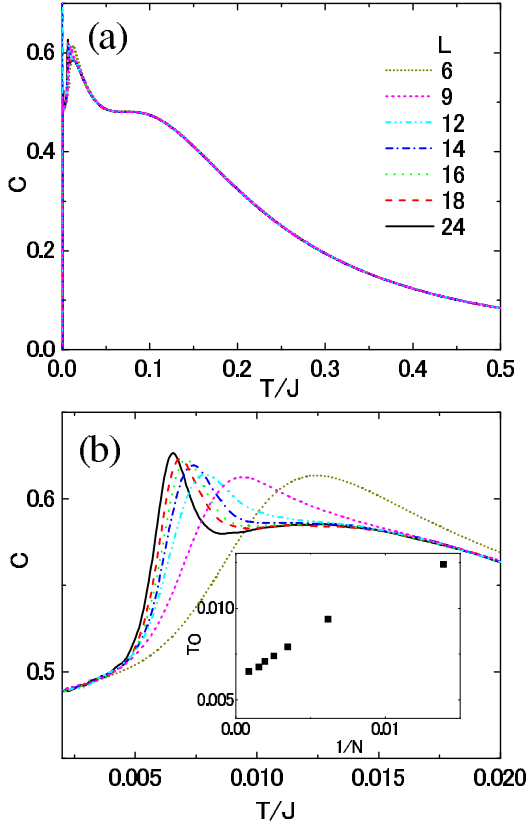}
\caption{ Specific heat calculated in several cluster sizes. 
Low temperature data are enlarged in (b). 
Inset in (b) shows a peak position $T_{\rm O}$ in the specific heat as function of $1/N$.} 
\label{fig:specific}
\end{figure}
\subsection{Monte Carlo simulation}

\begin{figure}
\includegraphics[width=0.8\columnwidth]{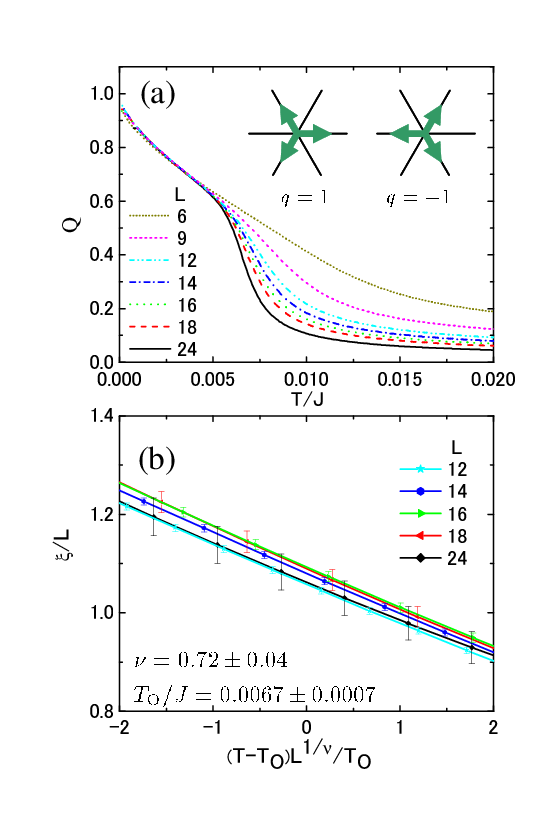}
\caption{ 
(a) Correlation function of a variable $q$ for the PS angle. 
Insets show the $q=1$ and $-1$ PS configurations. 
(b) Scaling analyses for the correlation length of $q_i=\cos 3 \theta_i$. 
} 
\label{fig:qqq}
\end{figure}

In the previous section, we assume the uniform PS configurations 
and show lifting of the continuous degeneracy by thermal fluctuation 
within the spin-wave scheme. 
Now we take off this restriction and show the results 
obtained by the MC simulation. 
Because of a limited system size in the MC calculation, 
both the spin wave and MC methods provide us complemented information with each other. 
To avoid a trap of the simulation in local minima,  
we adopt the multi-canonical MC technique. 
The energy distribution functions are obtained by the histogram method~\cite{berg} 
and the CFP one~\cite{lopes}. 
In most of the simulation, 
1$\times$10$^7$ MC steps are used to produce the energy histogram, 
and 2$\times$10$^8$ MC steps are for the calculation. 
Statistical averages and errors are obtained by 20 times simulations. 
Except for the results in Fig.~\ref{fig:qqq}(b), 
error bars are enough small and are not plotted in the figures. 
We adopt a cluster of $2 \times L \times L (\equiv N)$ sites with $L = 2 \sim 24$. 

First we present, in Fig.~\ref{fig:specific}, 
temperature dependence of the specific heat $C(T)$ 
for several system sizes. 
As seen in Fig.~\ref{fig:specific}(a), over all behavior 
does not show size dependence. 
There is a shoulder around $0.1J$ 
and a sharp peak around $0.005J-0.01J$ which depends on system size. 
Result in a $2 \times 5 \times 7$ size cluster is almost identical with that  
in $2 \times 6 \times 6$; a shape of the cluster is not essential. 
A magnification of $C(T)$ in a low temperature region 
is presented in Fig.~\ref{fig:specific}(b). 
With increasing a system size, 
the peak shifts to a lower temperature side and becomes sharp. 
The peak position is denoted as $T_{\rm O}$ from now on. 
As shown in the inset of Fig.~\ref{fig:specific}(b), 
$T_{\rm O}$ approaches a finite value about 0.0064$J$ in the thermodynamic limit. 
It is worth noting that this value of $T_{\rm O}$ is much smaller than 
the mean-field ordering temperature $3J/8$. 
At zero temperature limit, $C(T)$ takes about $0.5$ corresponding 
to one degree of freedom per site, i.e. the two-dimensional PS angle.

\begin{figure}
\includegraphics[width=0.8\columnwidth]{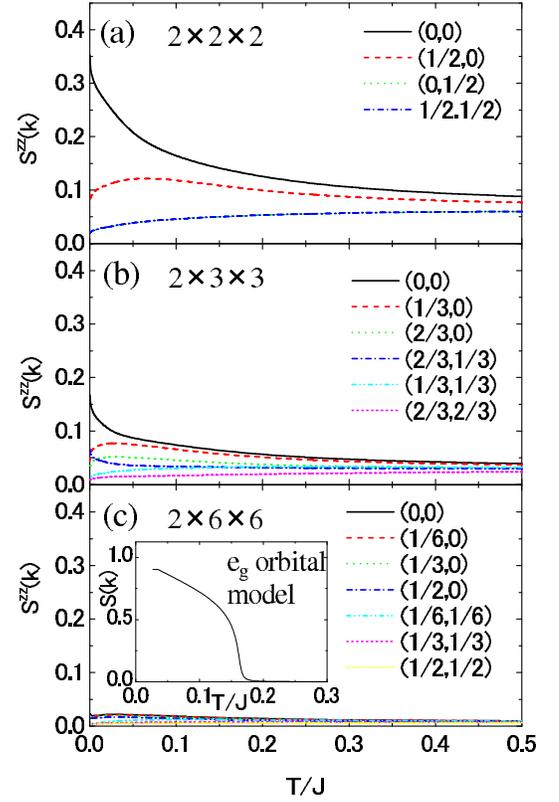}
\caption{ Correlation functions $S^{zz}({\bf k})$ for several momentum ${\bf q}$. 
Cluster sizes are $2 \times 2 \times 2$ in (a), $2 \times 3 \times 3$ in (b), and $2 \times 6 \times 6$ in (c).
Inset in (c) shows correlation function 
$S({\bf k})=4N^{-2} \sum_{ij} \langle {\bf T}_i \cdot {\bf T}_j  \rangle e^{i {\bf k} \cdot ({\bf r}_i-{\bf r}_j)}$ 
at ${\bf k}=(\pi, \pi, \pi)$ calculated in the $e_g$ orbital model. A cubic cluster with 18$^3$ sites is adopted.~\cite{tanaka05,tanaka08}} 
\label{fig:corr_class}
\end{figure}
To elucidate the PS structure below $T_{\rm O}$, 
we calculate the correlation functions for PS defined by 
\begin{eqnarray}
S^{lm}({\bf k})= \frac{4}{N^2}\sum_{ij} \langle T_i^l T_j^m \rangle 
e^{i {\bf k} \cdot \left ( {\bf r}_i-{\bf r}_j \right )} , 
\label{eq:correlation}
\end{eqnarray}
where $l$ and $m$ take $x$ and $z$, 
and ${\bf r}_i$ is a position of site $i$. 
The maximum value of the functions is one. 
The $z$ component of the correlation functions $S^{zz}(\bf k)$ 
for several system sizes are presented in Fig.~\ref{fig:corr_class}. 
We calculate $S^{zz}({\bf k})$'s for all possible momenta ${\bf k}$ in a cluster. 
In a $2 \times 2 \times 2$ cluster, $S^{zz}({\bf k})$ at ${\bf k}=(0,0)$ takes about 0.3 in low temperatures. 
However, with increasing $N$, the values of $S^{lm}({\bf k})$ decrease rapidly, 
and in a $2 \times 6 \times 6$ cluster, all $S^{lm}({\bf k})$'s are less than 3$\%$ of their maximum value. 
Other components, $S^{xx}({\bf k})$ and $S^{xz}({\bf k})$, are similar to $S^{zz}({\bf k})$. 
We conclude that, below $T_{\rm O}$, there are no conventional long-range order corresponding 
to the correlation functions given in Eq.~(\ref{eq:correlation}).  
This is not trivial for the present model where the Mermin-Wagner's theorem is not applicable. 
The present results are in contrast to those in the $e_g$ orbital model; 
the PS correlation function at ${\bf k}=(\pi, \pi, \pi)$  
starts to increase around $0.17J$, and approaches its maximum value at the low temperature limit, 
as shown in the inset of Fig.~\ref{fig:corr_class}(c).  

\begin{figure}
\includegraphics[width=0.8\columnwidth]{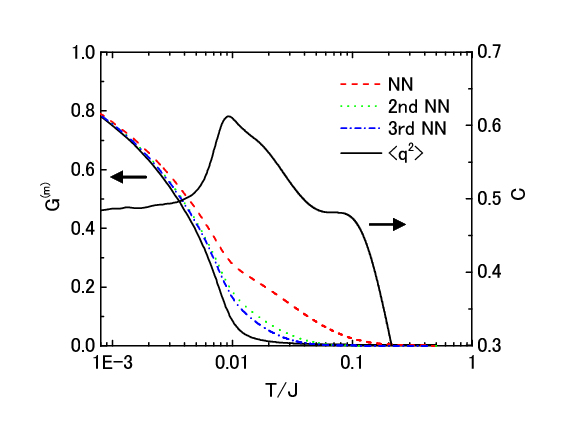}
\caption{ Temperature dependence of correlation functions of $q_i$ and specific heat.
Broken, dotted and dash-dotted lines are for the correlation functions between the NN, 2nd NN and 3rd NN sites, respectively.} 
\label{fig:short}
\end{figure}

Here we propose a physical parameter $q$ for the PS angle $\theta_i$ 
defined by 
\begin{eqnarray}
q=\frac{1}{N} \sum_i \cos 3 \theta_i . 
\label{eq:3theta}
\end{eqnarray}
Because the Hamiltonian is invariant under the inversion of all Ps's, 
$ \langle q \rangle $ is zero in a disordered phase. 
When the angle $\theta_i$ takes one of the three angles $2n \pi/3$ [$(2n+1)\pi/3$], 
$q=1 \ (-1)$ [see inset of Fig.~\ref{fig:qqq}(a)]. 
In Fig.~\ref{fig:qqq}, 
we plot the temperature dependence of the correlation function of $q$ defined by 
\begin{equation}
Q=\sqrt{\langle q^2 \rangle } . 
\label{eq:largeq}
\end{equation}
This starts to increase around $T_{\rm O}$ 
and is saturated to the maximum value at the low temperature limit. 
With increasing the system size $N$, 
$Q$ abruptly increases around $T_{\rm O}$.  
We show, in Fig.~\ref{fig:qqq}(b), the finite-size scaling plot for the correlation length of 
$q_i \equiv \cos 3 \theta_i$ defined by~\cite{lee06} 
\begin{equation}
\xi=-\frac{1}{G(0)} \frac{dG({\bf k})}{d|{\bf k}|^2} \bigg |_{{|\bf k}|=0} , 
\end{equation}
where we define the correlation function of $q_i$ as 
\begin{equation}
G({\bf k})=\frac{1}{N^2} \sum_{ij} \langle q_i q_j \rangle e^{i {\bf k}\cdot ({\bf r}_i-{\bf r}_j) } . 
\end{equation}
As shown in this figure, $\xi/L$ 
in several system sizes are scaled by the scaling function 
$(T-T_{\rm O}) L^{1/\nu}/T_{\rm O}$ within error bars. Here we obtain 
the transition temperature $T_{\rm O}=0.0067\pm 0.0007$ and $\nu=0.72 \pm 0,04$. 
These results imply that, at low temperature below $T_{\rm O}$,  
the PS angle at each site takes one of the three angles $2n\pi/3$ ($\cos 3 \theta=1$),  
or one of $(2n+1)\pi/3$ ($\cos 3 \theta=-1$). 
When the $q_i=1$ and $-1$ states are randomly distributed in a lattice, 
the correlation function $Q$ should be zero. 
It is found from the snapshot of the MC simulation 
that the three $q_i=$1 states, or the three $-1$ states, coexist below $T_{\rm O}$. 
From the view point of the PS angle, 
a shoulder structure in $C(T)$ around $T/J=0.1$ shown in Fig.~\ref{fig:specific} 
corresponds to development of the short range correlation. 
In Fig.~\ref{fig:short}, we show the short-range correlation functions of $q_i$ defined by 
\begin{equation}
G^{(m)}=\frac{1}{z^{(m)}N} \sum_{(ij)}^\prime \langle q_i q_j \rangle ,  
\end{equation}
where $G^{(m)}$ with $m=$1, 2 and 3 are the correlations between NN, the next NN and the 3rd NN sites, respectively. 
A numerical factor $z^{(m)}$ is a number of the neighboring pairs, and $\sum_{(ij)}^\prime$ 
represents a sum of the pairs. 
It is clearly shown in Fig.~\ref{fig:short} that a shoulder of $C(T)$ corresponds 
to development of $G^{(1)}$. 

\begin{figure}
\includegraphics[width=0.8\columnwidth]{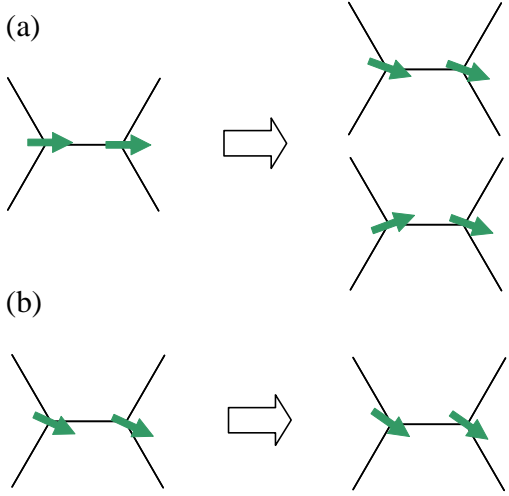}
\caption{ 
Schematic view of the $q=1$ and $q \ne 1$ states and their low-lying fluctuations. 
} 
\label{fig:fluctuation}
\end{figure}

Stability of the $q=\pm 1$ states is attributed to the low-lying excited states around the $q=\pm 1$ states. 
Consider one of the $q=1$ states shown in Fig.~\ref{fig:fluctuation}(a), 
and local PS fluctuations in a certain NN bond in this state. 
There are two ways for fluctuation where the condition in Eq.~(\ref{eq:condition}) is satisfied in this bond. 
That is, the two kinds of excited states appear thermally with the same probability. 
Situation is different away from the $q = 1 $ configuration. 
Consider one of the $q \ne \pm1 $ states shown in Fig.~\ref{fig:fluctuation}(b). 
There is only one way for fluctuation where the condition in Eq.~(\ref{eq:condition}) is satisfied.  
This high density of the low-lying fluctuations around $q= \pm 1$ states 
contributes to the entropy gain and stabilizes the $q= \pm 1$ states at finite temperature.~\cite{wu_comment}

As shown above, we have found that, below $T_{\rm O}$, the PS angle at each site is fixed at one of the three angles $2n\pi/3$ or one of the three $(2n+1)\pi/3$. 
Within the present calculations, we do not insist whether 
all $q=\pm 1$ states are realized equivalently or not.  
We will discuss this point in Sect.~\ref{sect:conclusion} in more detail with supplementary calculations. 

\section{Quantum orbital state}
\label{sect:quantum}
In this section, we analyze the Hamiltonian in Eq.~(\ref{eq:ham}) 
where the PS is treated as a quantum spin operator with a magnitude of $S=1/2$. 

\subsection{Spin wave analyses}
\label{sect:sw_quantum}
\begin{figure}
\includegraphics[width=0.8\columnwidth]{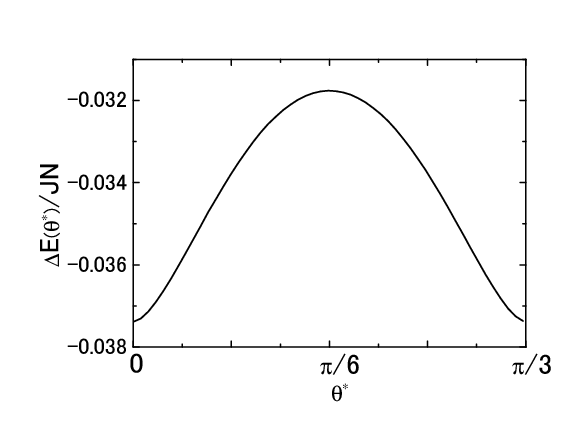}
\caption{ Energy correction $\Delta E(\theta^\ast)$ from the zero-point vibration as a function of the orbital angle $\theta^\ast$. } 
\label{fig:hp}
\end{figure}

To elucidate roles of the quantum fluctuation on the stable orbital state at zero temperature, 
we start from, for simplicity, the uniform orbital state with an PS angle $\theta^\ast$. 
The Hamiltonian is analyzed 
by using the Holstein-Primakoff transformation.~\cite{kubo02} 
We utilize the rotating frame given by the unitary transformation with respect to the $T^y$ axis, 
and introduce the two Holstein-Primakoff bosons, $a_{\bf k}$ and $b_{\bf k}$, 
for the two sublattices in the honeycomb lattice. 
The Hamiltonian up to the second order of the boson operator is given as 
\begin{eqnarray}
{\cal H}_{\rm SW}&=&-\frac{3}{4}S^2JN+\frac{3}{2}SJ \sum_{{\bf k}} 
\bigg [ a_{\bf k}^\dagger a_{\bf k}+b_{\bf k}^\dagger b_{\bf k}
\nonumber \\
&-&
\frac{1}{2} 
\bigg ( 
 \gamma_{\bf k} (\theta^\ast) a_{\bf k} b_{-\bf k}
+\gamma_{\bf -k}(\theta^\ast) a_{\bf k}^\dagger b_{-\bf k}^\dagger
\nonumber \\
&+&\gamma_{\bf k} (\theta^\ast) a_{\bf k} b_{\bf k}^\dagger
+\gamma_{\bf -k}(\theta^\ast) a_{\bf k}^\dagger b_{\bf k} \bigg ) \bigg ], 
\label{eq:sw1}
\end{eqnarray}
where $\gamma_{\bf k}(\theta^\ast)$ is the structure factor  
defined by 
\begin{equation}
\gamma_{\bf k}(\theta^\ast)=
\frac{2}{3} \sum_{\eta} \sin^2 \left ( \theta^\ast -\frac{2 \pi}{3} n_\eta \right) 
e^{-i {\bf k} \cdot {\bf e}_\eta}, 
\end{equation}
with a numerical factor $(n_\alpha, n_\beta, n_\gamma)=(0,1,2)$. 
The first term in Eq.~(\ref{eq:sw1}) is the zero-th order energy, denoted by $E_0$, 
which corresponds to the second term of Eq.~(\ref{eq:ham2}) in the classical spin case. 
This energy is  independent of the angle $\theta^\ast$, as mentioned previously. 
By applying the Bogoliubov transformation, 
we diagonalize the second term and obtain, 
\begin{eqnarray}
{\cal H}_{\rm sw}=E_0+\Delta E(\theta^\ast)+\sum_{{\bf k} \ l=\pm } 
\omega^{(l)}_{\bf k}(\theta^\ast) c^{(l) \dagger }_{\bf k} c^{(l)}_{\bf k}  , 
\label{eq:sw2}
\end{eqnarray}
where we introduce the boson (orbiton) operators $c^{(\pm)}_{\bf k}$ and 
their energy dispersions 
\begin{equation}
\omega^{(\pm)}_{\bf k}(\theta^\ast)=\frac{3}{2} SJ \sqrt{1 \pm |\gamma_{\bf k}(\theta^\ast)|}. 
\end{equation}
In the case of $\theta^\ast=n \pi/3$, 
these dispersions show an one-dimensional character; 
for example, $\omega^{(\pm)}_{\bf k}(\theta^\ast=0)=(3SJ/2) \sqrt{1 \pm \cos(k_ya/2)}$ 
which is independent of $k_x$, 
where we define $k_x={\bf k} \cdot {\bf G}_a$ and $k_y={\bf k} \cdot ({\bf G}_a+2{\bf G}_b)/\sqrt{3}$. 
The second term in Eq.~(\ref{eq:sw2}) corresponds to 
the correction from the zero-point vibration. 
This is given as 
\begin{eqnarray}
\Delta E(\theta^\ast)&=&  \frac{\sqrt{3}a^2N}{32 \pi^2} 
\int_{\rm 1stBZ} dk_x dk_y\nonumber \\
& \times& 
\left [ \omega^{(+)}_{\bf k}(\theta^\ast)+\omega^{(-)}_{\bf k}(\theta^\ast) -2 \right ] , 
\end{eqnarray}
where $a$ is the NN bond length, and $\int_{\rm 1stBZ} dk_xdk_y$ represents the integral in the 1st Brillouin zone. 
Numerical results of $\Delta E(\theta^\ast)$ as a function of $\theta^\ast$ are presented in Fig.~\ref{fig:hp}. 
The energy correction takes its minimum at six angles of $\theta^\ast=n \pi/3$ with integer number $n$, 
reflecting the C$_6$ symmetry in the free energy.~\cite{zhao_comment} 
It is worth noting that these are the same angle 
where the classical free energy takes the minimum (see Fig.~\ref{fig:f_theta}). 
That is, both the quantum and thermal fluctuations stabilize the same orbital configurations 
within the uniform PS alignments. 
Stability at these angles in the quantum model is attributed to the dispersion relation of the orbitons 
$\omega_{\bf k}^{(\pm)}(\theta^\ast)$; 
at $\theta^\ast=n \pi/3$, there is a one-dimensional 
zero-energy mode. 
For example, $\omega^{(-)}(\theta^\ast=0)=0$ along the $(k_x,k_y)=(0,0)$ to (1,0) direction. 
This low-lying excitation contributes to the energy gain from the quantum zero-point fluctuation. 
We suppose that, when the higher-order terms corresponding to the orbiton-obtion interactions, 
are taken into account, the dispersion becomes gap-full, and the energy gain due to the zero-point fluctuation 
is reduced. 

\begin{figure}
\includegraphics[width=0.75\columnwidth]{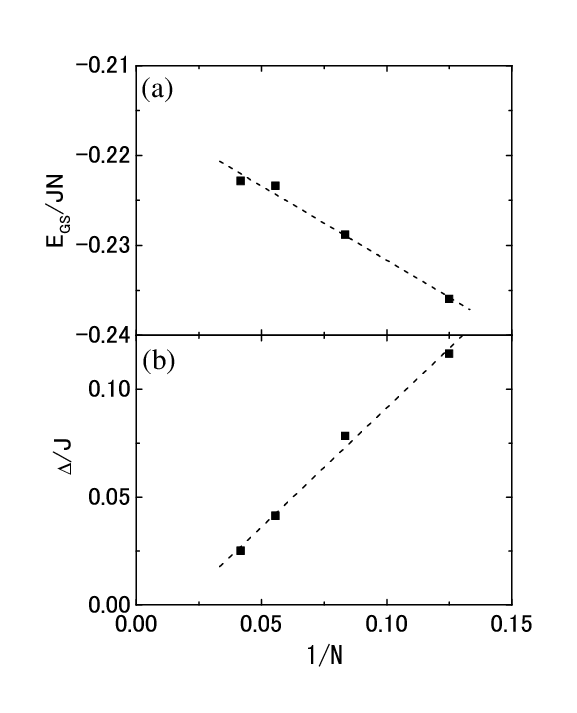}
\caption{ (a) Ground state energy and (b) energy gap for several size clusters. 
For the $2 \times 3 \times 3$ cluster, 
energy difference between the doubly degenerate ground states and the first excited states are plotted.
Broken lines are obtained by the least square fittings. 
}
\label{fig:gap}
\end{figure}

\subsection{Lanczos method}
\label{sect:lanczos}

\begin{figure}
\includegraphics[width=0.8\columnwidth]{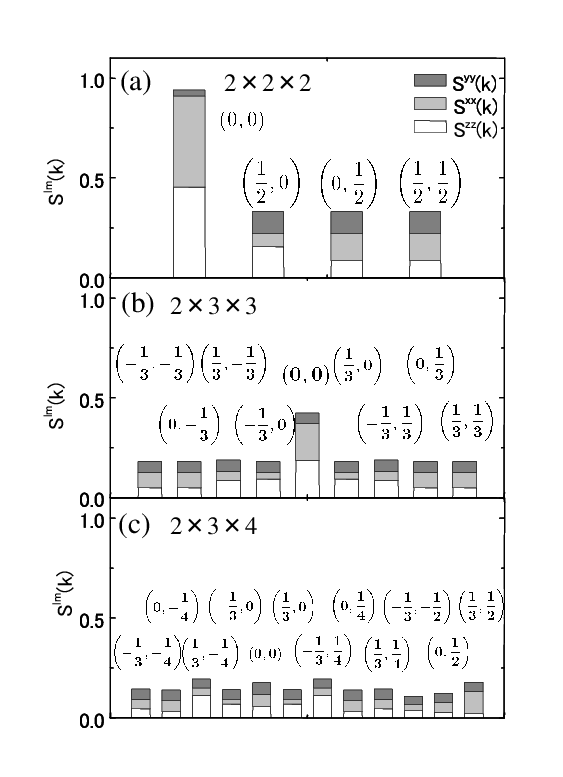}
\caption{Correlation functions $S^{lm}({\bf k})$ for several momenta ${\bf k}$. 
Cluster sizes are $2 \times 2 \times 2$ in (a), $2 \times 3 \times 3$ in (b), and $2 \times 3 \times 4$ in (c).
} 
\label{fig:corre_quantum}
\end{figure}

In the numerical calculation in a finite size, 
the orbital structure can be examined without assumption 
of the uniform PS configurations. 
In the quantum Monte-Calro simulation, we met a serious negative sign problem. 
Here we use the exact diagonalization technique based on the Lanczos algorithm. 
We adopt finite size clusters of $N=2 \times 2 \times 2$, $2 \times 2 \times 3$, 
$2 \times 3 \times 3$, and $2\times 3 \times 4$ sites 
with the periodic boundary condition. 
Because of no conserved quantities in the Hamiltonian, 
all state vectors in the Hilbert space of $2^N$ dimension are dealt with 
in the Lanczos calculation. 

First, we show the ground state energy $E_{\rm GS}$ and the energy gap $\Delta$ for several size clusters in Fig.~\ref{fig:gap}. 
The ground state energy tends to approach, in the thermodynamic limit, 
around $-0.215NJ$ which is higher a little than 
the spin-wave results $E_0+\Delta E(\theta^\ast)=-0.225NJ$ at $\theta^\ast=n \pi/3$. 
Except for the $2 \times 3\times 3$ cluster, 
the ground state is not degenerate. 
The gap energy is defined as an energy difference between the ground state and the 1st excited one. 
The numerical value monotonically decreases with the system size $N$, 
and seems to vanish in the thermodynamic limit. 
However, we cannot distinguish 
the two possibilities in an infinite system: 
degenerate ground states and a non-degenerate one with gap-less excitation. 
The correlation functions of PS defined in Eq.~(\ref{eq:correlation}) 
are calculated for several momenta and system sizes (Fig.~\ref{fig:corre_quantum}). 
In the smallest size of $2 \times 2 \times 2$ sites, 
$S^{zz(xx)}({\bf k})$ at ${\bf k}=(0,0)$ stands out. 
However, with increasing $N$, 
$S^{lm}({\bf k})$'s become almost momentum independent 
and all of the values are less than 25$\%$ of the maximum. 
Reduction of $S^{lm}({\bf k})$ at ${\bf k}=(0,0)$ 
is faster than $1/N$. 
Thus, the conventional long-range order 
characterized by the correlation functions does not exist, 
as we have shown in the classical model. 

\begin{figure}
\includegraphics[width=0.9\columnwidth]{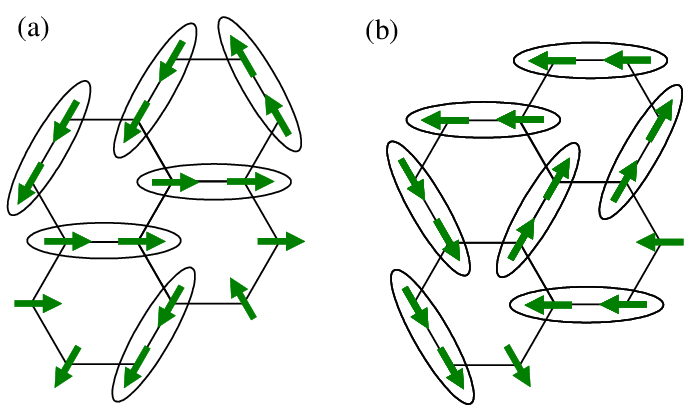}
\caption{ 
Some of the PS configurations where 
the honeycomb lattice is covered by NN bonds with the minimum bond energy. 
One of the $q=1$ states in (a),  and one of the $q=-1$ in (b). 
In NN bonds surrounded by ellipses, the bond energy is the lowest. 
}
\label{fig:cover}
\end{figure}
\begin{figure}
\includegraphics[width=0.9\columnwidth]{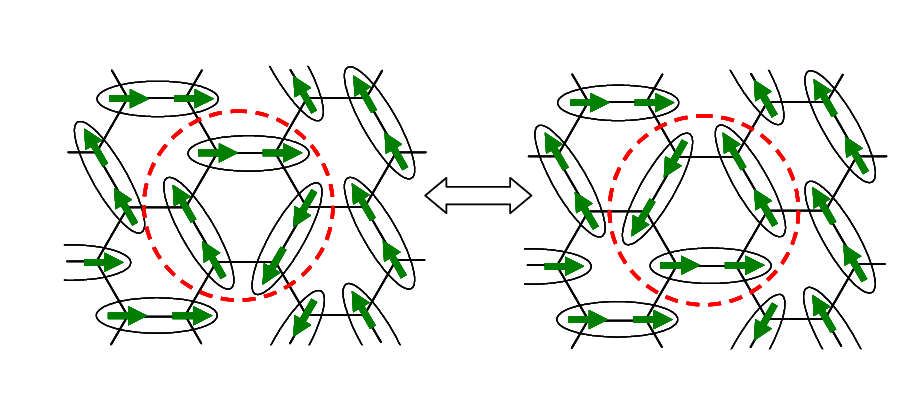}
\caption{ 
One example for the two PS configurations where a resonance state is possible due to the off-diagonal matrix element.  
}
\label{fig:resonance}
\end{figure}

\begin{figure}
\includegraphics[width=0.8\columnwidth]{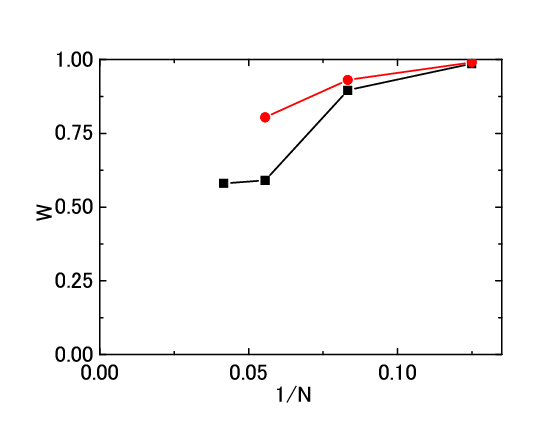}
\caption{ Overlap integrals between the ground-state wave function and the trial functions. 
Squares are for the results where the variational parameters ${\cal A}_l$ are taken to be one, 
and circles are obtained by optimizing ${\cal A}_l$. 
}
\label{fig:weight}
\end{figure}

In the quantum system, 
the operator corresponding to $q=N^{-1}\sum_i \cos 3 \theta_i$ defined in Eq.~(\ref{fig:qqq}) 
becomes a constant $C$-number due to the algebra for the $S=1/2$ spin operator. 
Then, we adopt the variational-like method to analyze the ground-state wave function. 
As explained in Sect.~\ref{sect:classical}, 
the classical PS states below $T_{\rm O}$ is characterized by the parameter $q$ 
defined in Eq.~(\ref{eq:3theta}), 
i.e. the PS angles are fixed at $2n\pi/3$ or $(2n+1)\pi/3$ with an integer number $n$. 
From these results and snapshots of the MC simulation, 
we consider the trial PS states where 
a honeycomb lattice is covered by the NN bonds with the minimum bond energy. 
Some examples are shown in Fig.~\ref{fig:cover}. 
We construct the wave function as a linear combination of these states. 
This is given by  
\begin{eqnarray}
|\Psi^{(\pm)} \rangle={\cal N} \sum_l {\cal A}_l  
\left \{ |\psi_l^{(\uparrow)} \rangle \pm |\psi_l^{(\downarrow)} \rangle \right \}, 
\label{eq:wave}
\end{eqnarray}
where ${\cal N}$ is a normalized factor, ${\cal A}_l$ are variational parameters, 
and $|\psi_l^{(\uparrow,\downarrow)} \rangle$ is the wave function for the $l$-th PS configuration 
which satisfy the condition explained above. 
The wave function $|\psi_l^{(\uparrow)} \rangle $ is given by the unitary transformation 
from the all-up PS state $| \uparrow \cdots \uparrow \rangle $  
as follows, 
\begin{eqnarray}
|\psi_l^{(\uparrow)} \rangle =\prod_{\langle ij \rangle_l} U(\phi_\eta)_{\langle ij \rangle_l}
| \uparrow \cdots \uparrow \rangle . 
\end{eqnarray}
Similarly, $|\psi_l^{(\downarrow)} \rangle $ is obtained from 
the all-down state $| \downarrow \cdots \downarrow \rangle $. 
The $\alpha$ bond direction is taken as the quantized axis, and 
the subscript $\langle ij \rangle_l$ represents the NN $ij$ pair in the $l$-the PS configuration. 
The operator $U(\phi_\eta)_{\langle ij \rangle_l}$ describes a rotation of ${\bf T}_i$ and ${\bf T}_j$ 
with respect to the $T^y$ axis 
defined by 
\begin{eqnarray}
U(\phi_\eta)_{\langle ij \rangle_l}=\exp \left[-i\phi_\eta \left(T_i^y+T_j^y \right) \right], 
\end{eqnarray}
where $\eta$ indicates a direction connecting $i$ and $j$, 
and $(\phi_\alpha, \phi_\beta, \phi_\gamma)=(0, 2\pi/3, 4\pi/3)$. 
Because of the off-diagonal matrix elements 
among some states in $|\psi_l^{(\uparrow)} \rangle $ and $|\psi_l^{(\downarrow)} \rangle $, 
certain kinds of resonance states are realized. 
A set of two PS configurations, termed $|\psi_L \rangle $ 
and $| \psi_R \rangle$, shown in Fig.~\ref{fig:resonance} is an example.  
The off-diagonal matrix element between the two is  
$\langle \psi_L|{\cal H}_J|\psi_R \rangle=-JN3/(16 \cdot 2^6)$. 
This is about 10$\%$ of the energy gain due to quantum effect, 
$(E_{\rm GS}-E_{0})/JN$, where $E_{\rm GS}$ is the ground state energy shown in Fig.~\ref{fig:gap}(a) and $E_0=-3JN/16$. 
Since the Hamiltonian is invariant under the inversion of all PS operators, 
the energy eigen states are classified by the parity of this operation. 
The wave functions $|\Psi^{(+)} \rangle$ and $|\Psi^{(-)} \rangle$ have the even and odd parities, respectively.  
Except for the degenerate ground state in the $2 \times 3 \times 3$ size cluster, 
the ground state wave function $|0 \rangle$ shows the even parity. 
The doubly degenerate ground states in $2 \times 3 \times 3$ are classified as  
the even and odd parity states, and the even-parity one is used for the analyses. 
Figure~\ref{fig:weight} shows the overlap integral $W \equiv |\langle 0 | \Psi^{(+)} \rangle|^2$ 
as a function of $1/N$. 
In a $2 \times 2 \times 2$ size cluster, the ground-state wave function is almost completely reproduced by 
the trial function. 
With increasing $N$,  
a value of $W$ is gradually reduced. 
However, this reduction is rather weak by optimizing the variational parameters ${\cal A}_l$, 
and $W$ is maintained around 0.8 even in the largest size cluster. 
Thus, at least within the present calculation, 
the ground-state wave function is well reproduced by the trial wave function 
where the honeycomb lattice is covered by NN bonds with the minimum bond energy. 

\section{discussion and summary}
\label{sect:conclusion}

First, we have some remarks on low temperature orbital state in the classical model. 
As shown in Sect.~\ref{sect:classical}, below $T_{\rm O}$, 
the PS angle at each site is fixed at one of the three angles $2n\pi/3$ or one of the three $(2n+1)\pi/3$.  
Here we discuss whether all $q = \pm 1$ states appear equivalently or 
some specific PS configurations in the $q = \pm 1$ states are stabilized.  
First we are able to exclude a possibility of the so-called directional order (DO). 
This is well examined in the orbital compass model in a two-dimensional square lattice;~\cite{mishra03,nussinov05,tanaka07}
one component in the PS operator, e.g. $T^z$, is aligned uniformly in each one-dimensional chain along 
a direction in the square lattice, e.g. the $z$ direction, but there is no PS correlation between the different chains. 
A natural order parameter of DO is 
$D_{\rm compass}=\sum_i \left (  T_{i}^zT_{i+{\bf e}_z}^z-T_{i}^xT_{i+{\bf e}_x}^x \right ) .$
Below the DO temperature, 
a PS-angle parameter $\sum_i \cos 2 \theta_i$, such as $q$ in Eq.~(\ref{eq:3theta}), is developed, 
but the conventional PS correlation functions, such as $S^{lm}({\bf k})$ in Eq.~(\ref{eq:correlation}), are not.   
We introduce the honeycomb lattice version of the directional order parameter: 
\begin{equation}
D=\sum_{i \in A \ \eta} \tau_i^\eta \tau_{i + {\bf e}_\eta}^\eta e^{i 2 n_{\eta} \pi/3}  . 
\end{equation}
When PS's are aligned uniformly inside zigzag chains along $\eta$ direction, but these chains are independent with each other, $D$ acts as a monitor. 
However, calculated $\langle D^2 \rangle^{1/2}$ by the MC method 
are less than $5 \times 10^{-4}$ and quickly disappears with increasing the system size. 
We also consider that a possibility of the Kosterlitz-Thouless transition~\cite{tobochnik79,faloi89} at $T_{\rm O}$ is weak. 
We calculate the uniform susceptibility 
\begin{equation}
\chi_u=\frac{4}{TN} \sum_{ij} \langle {\bf T}_i \cdot {\bf T}_j \rangle, 
\end{equation}
and the corresponding correlation length $\xi_u$. 
Though the present system size is limited up to $2 \times 24 \times 24$ sites, 
both $\chi_u$ and $\xi_u$ do not show anomalous behavior around $T_{\rm O}$,  and 
their values decrease with increasing the system size. 

\begin{figure}
\includegraphics[width=0.8\columnwidth]{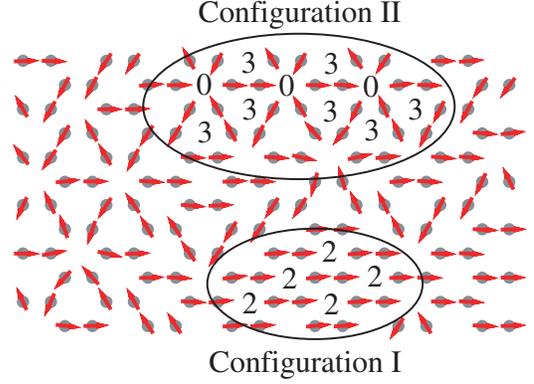}
\caption{ A MC snapshot for PS configurations. 
Numbers in each hexagons denotes a number of the minimum energy bonds $n_{\rm min}$. 
}
\label{fig:snap}
\end{figure}

\begin{figure}
\includegraphics[width=0.85\columnwidth]{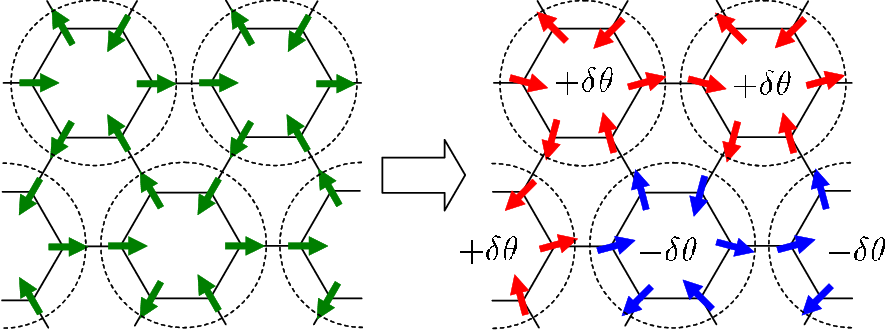}
\caption{ 
Zero-dimensional fluctuation in the configuration II. 
Left: PS configuration. Right: 
a configuration obtained by $\pm \delta \theta$ rotation of PS's in each hexagon with $n_{\rm min}=0$. 
}
\label{fig:hexagon}
\end{figure}

Here, we suggest that some topological PS configurations in hexagons may be more stabilized 
than other $q=\pm 1$ states in the classical model. 
In a MC snapshot (see Fig.~\ref{fig:snap}), 
we often find two characteristic PS configuration patterns in a honeycomb lattice; 
a uniform PS array termed configuration I, 
and a regular array of hexagons with maximum energy gain, termed configuration II. 
Of course, their simple long-range orders are excluded 
from the calculated results of $S^{lm}({\bf k})$ in Fig.~\ref{fig:corr_class}. 
There are possibilities that two configurations coexist, 
and/or these are distributed randomly. 
These are monitored by a parameter $n_{\rm min}$ 
which represents a number of NN bonds with the minimum bond energy in a hexagon. 
In the configuration II, hexagons with $n_{\rm min}=3$ and 0 are aligned regularly. 
It is convenient to introduce the following parameter defined in a hexagon at ${\bf r}$; 
\begin{equation}
R({\bf r})=\frac{9}{8} \left \{ \frac{2}{3} N({\bf r})-1 \right \}^2-\frac{1}{8} , 
\end{equation} 
with 
\begin{equation}
N({\bf r})= \sum_{(ij)} \left ( \frac{16}{3} \tau_i^{\eta_{ij}} \tau_j^{\eta_{ij}} -\frac{1}{3} \right )  , 
\end{equation}
where $\sum_{(ij)}$ represents a sum for six NN bonds in a hexagon.  
Because $\langle \tau_i^{\eta_{ij}} \tau_j^{\eta_{ij}} \rangle=1/4$ 
when a NN $ij$ bond takes the minimum bond energy, 
we have $\langle N({\bf r}) \rangle =n_{\rm min}$. 
The parameter $R({\bf r})$ 
takes one for the hexagons with $n_{\rm min}=0,3$, and zero for the hexagon with $n_{\rm min}=1,2$. 
We calculate $N^{-1} \langle \sum_{\bf r} R({\bf r}) \rangle$ in the $2 \times 9 \times 9$ 
site cluster by the MC method. 
The calculated value is about 0.42 below $T_{\rm O}$ which is larger  
than a value $(0.3)$ in the states where all $q = \pm 1$ configurations appear equivalently. 
That is, the configuration II is expected to be more stabilized than other $q=\pm 1$ states. 
This is due to their low-energy fluctuations. 
Hexagons characterized as $n_{\rm min}=0$ are included in the configuration II. 
As shown in Fig.~\ref{fig:hexagon}, there are two-ways of fluctuation in each hexagon with $n_{\rm min}=0$.  
When we consider the configuration II containing $m$ hexagons with $n_{\rm min}=0$, 
a number of configurations are roughly $ _{N/6}C_{m} 2^m$. 
This is remarkable in comparison with that in the configuration I;  
as explained in Fig.~\ref{fig:theta0}, there are also two ways of fluctuation in each zigzag chain. 
This corresponds to the so-called stacking degeneracy observed in the $e_g$ orbital model.~\cite{nussinov04} 
When we consider the configuration I containing $m$ zigzag chains, 
a number of configurations are roughly $ _{\sqrt{N}}C_{m} 2^m$. 
Difference between the two is attributed to dimensionality of the fluctuations. 
This zero-dimensional fluctuation seen in the configuration II is unique in this honeycomb lattice model, 
and is expected to be an origin of no conventional long-range order. 

In summary, we study the doubly degenerate orbital model on a honeycomb lattice, 
motivated from an orbital state in multiferroic layered iron oxides $R$Fe$_2$O$_4$. 
There is a macroscopic number of degeneracy in the classical ground state, 
as seen in the three-dimensional $e_g$ orbital model. 
We mainly focus on lifting of the degeneracy due to thermal and quantum effects. 
In the classical and quantum spin-wave analyses, 
where the uniform orbital configurations are assumed, results are similar to  
those in the $e_g$-orbital model. 
Both thermal and quantum fluctuations stabilize the states 
with the PS angles of $\theta^\ast=n \pi/3$. 
Beyond the uniform configuration assumption, 
we apply the Monte-Carlo simulation to the classical model. 
A peak structure in the specific heat is found around $T_{\rm O}/J=0.006$. 
However, below $T_{\rm O}$, the PS correlation functions indexed by any possible momenta 
in clusters are not developed, unlike those in the $e_g$-orbital model. 
We find that the correlation function 
of a parameter for the orbital PS angle, $q=\sum_i \cos 3 \theta_i$,  
grows up below $T_{\rm O}$, and reaches its maximum at the low temperature limit. 
That is, the PS angle at each site takes one of the three angles $2n\pi/3$ 
or one of the three $2(n+1)\pi/3$. 
This degeneracy lifting is attributed to existence of low-lying fluctuation around these configurations. 
We suggest that zero-dimensional fluctuation 
in the hexagons plays a crucial role to make difference between the present honeycomb-lattice model and the 
$e_g$ orbital one in a cubic lattice.
We also analyze the quantum model by utilizing the Lanczos method. 
As seen in the classical model, 
any remarkable features are not shown in the two-body PS correlation functions. 
This suggests no conventional long-range order indexed by specific momenta. 
The ground-state wave function is analyzed by a variational-like method. 
This is well represented by a linear combination of the wave functions 
where a honeycomb lattice is covered by NN dimers with the minimum-energy PS configurations. 

\begin{acknowledgments}
The authors would like to thank 
M.~Sasaki, M. Matsumoto, H. Matsueda and T.~Tanaka for their valuable discussions. 
This work was supported by JSPS KAKENHI (16104005), and 
TOKUTEI 
``High Field Spin Science in 100T'' (18044001),  
``Novel States of Matter Induced by Frustration'' (19052001), and 
``Invention of Anomalous Quantum Materials''(19014003) from MEXT, 
NAREGI, and CREST. 
\end{acknowledgments}
$^{\ast}$ Present address: Japan Medical Materials Co., Osaka, 532-0003 Japan. 

\begin{references}
\bibitem{wu_comment}
These results imply that free hexagon loops in the honeycomb lattice suggested in 
Ref.~\onlinecite{wu08} are unstable in finite temperatures, since the PS angles of $(2n+1)\pi/6$ are required 
to make the non-interacting NN bonds.
\bibitem{zhao_comment}
A similar calculation was presented in Ref.~\onlinecite{zhao08}. 
However, the calculated energy does not show the C$_{6}$ symmetry 
in the PS angle space.
\end{references}
\end{document}